# Growth of immobilized DNA by polymerase: bridging nanoelectrodes with individual dsDNA molecules

Veikko Linko,*[a] Jenni Leppiniemi,[b] Boxuan Shen,[a] Einari Niskanen,[c] Vesa P. Hytönen[b] and J. Jussi Toppari[a]

We present a method for controlled connection of gold electrodes with dsDNA molecules (locally on a chip) by utilizing polymerase to elongate single-stranded DNA primers attached to the electrodes. Thiol-modified oligonucleotides are directed and immobilized to nanoscale electrodes by means of dielectrophoretic trapping, and extended in a procedure mimicking PCR, finally forming a complete dsDNA molecule bridging the gap between the electrodes. The technique opens up opportunities for building from the bottom-up, for detection and sensing applications, and also for molecular electronics.

## 1. Introduction

During recent years DNA has been proven to be a very flexible and versatile molecule within almost any areas of nanotechnology.[1] Many DNA-based templates and constructs,[2-4] even functional ones,[5] have been introduced as well as numerous assays making use of the unique properties of DNA.[6,7] Various approaches to implement molecular devices with a broad range of functions for nanoelectronics and bionanotechnology, are mostly based on superior self-assembly properties of DNA and well-developed tool kits, including many manipulation techniques. However, precise spatial control of diverse molecular components and assemblies is still a huge challenge.

One of the methods which has already enriched the possibilities of DNA as a key player in bionanotechnology is Polymerase Chain Reaction (PCR). It is one of the basic tools in molecular biology and the principle of the reaction was already invented in 1970,[8] but was not implemented to the laboratory work until during 1990's – Nobel Prize was awarded to Kary B. Mullis in 1993. PCR allows exponential amplification of a target DNA sequence, a *template*, by using short synthetic DNA oligonucleotides as reaction *primers*. The method employs thermostable polymerase and temperature cycling, which enables repetition of the amplification step several times, thus allowing one to detect and characterize even a single copy of DNA. There exist countless applications for PCR in fields such as diagnostics, juridical research and personal medicine, just to mention a few.

Typically, a PCR reaction is performed in solution. However, it has been shown that PCR can also be carried out when a DNA primer used for amplification is immobilized on a substrate. The first demonstration of "immobilized PCR" was carried out by Rasmussen et al. (1994),[9] who successfully detected leukemia virus and Salmonella by utilizing PCR with one of the primers covalently immobilized via 5'-phosphate group by carbodiimide chemistry. Several reports have applied this principal methodology to develop novel DNA-based methods, such as preparation of arrays of long DNA sequences by "on-chip" elongation.[10] A method, where both of the primers are immobilized in large amounts to extensive surfaces, has also been demonstrated, named as "bridge amplification".[11]

Here, we introduce a concept of growing individual dsDNA molecules locally on a chip from immobilized primers with nanoscale precision by mimicking a typical PCR procedure (however, we are not amplifying any target DNA sequence). The presented technique can serve as a bridge between novel DNA-nanotechnology and common molecular biology methods, while resolving also the problem of the precise spatial control.

The method is based on directed concentration and immobilization of short thiol-modified single-stranded primers to the ends of fingertip-type gold nanoelectrodes by utilizing alternating-current dielectrophoresis (AC-DEP) (see Fig. 1(a)). Dielectrophoresis consists in the movement of any polarizable particle in a nonuniform electric field,[12] and it has been widely exploited in trapping of diverse variety of objects,[13,14] including controllable micron-[15-17] and nanoscale[17-22] manipulation and positioning of DNA. The DEP-immobilized primers are then extended by polymerase during repetitive cycles similar to the PCR. During the annealing steps the elongated strands can *1)* form a complete dsDNA molecule having a template sequence and bridging the gap between the electrodes, or *2)* pair with the complementary strands originated from a template (Figs. 1(b), 3(e)-(h)). Yet, by utilizing multielectrode geometries, it is possible to immobilize primers only to the desired electrodes (see Fig. 1(a)) enabling specific and sequence depended growing of dsDNA between the chosen electrodes. This provides new possibilities for building from the bottom-up with DNA, and the method can find applications in the fields of detecting and sensing as well as in molecular electronics.

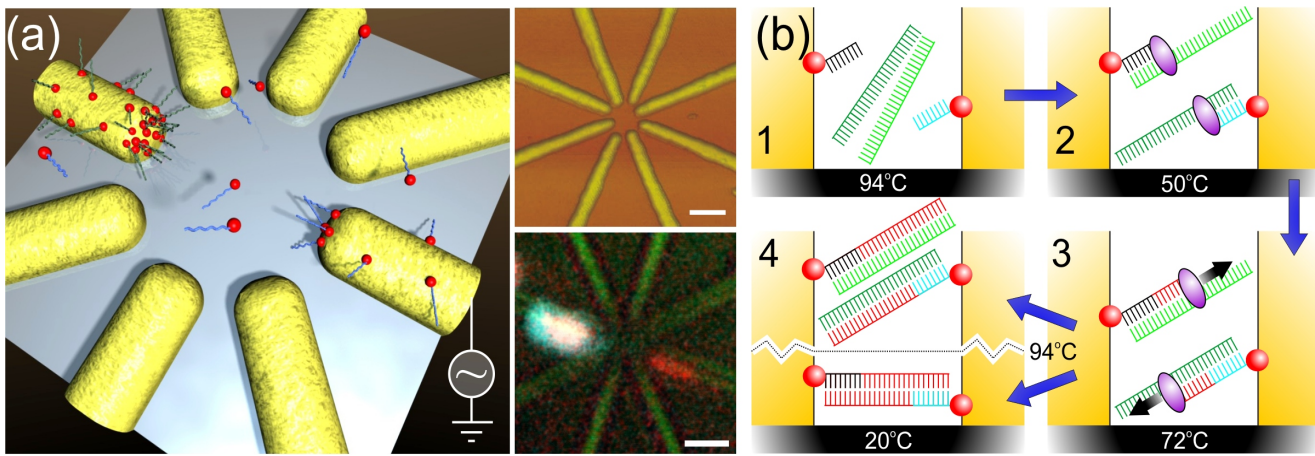

**Fig. 1** (a) *Left:* A schematic view of the electrode-specific dielectrophoretic (DEP) trapping of different thiol-modified primers with a multielectrode system. An AC voltage signal is applied one by one to the desired electrode gathering the primers to its end (also along the whole electrode), while keeping the other electrodes grounded. In the figure the electrode connected to an AC voltage source is collecting primers (blue tail), while the immobilization of other type of primers (green tail) to the opposite electrode has already been performed. *Upper right:* An AFM image of the multielectrode structure. The scale bar is 100 nm. *Lower right:* A confocal microscope image of 40 nt long primers labeled with different dye molecules [Cy3 (red) and Cy5 (blue)] separately trapped and immobilized to the opposite electrodes. The scale bar is 1 μm. The multielectrode geometry shown here was not utilized in the actual elongation experiments, but was used to show that the electrode-specific trapping is achievable enabling sophisticated wiring systems through the further optimization. (b) A sequence of schematic cross-sectional images presenting the polymerase growing method. A sample, containing primers (black and light blue) immobilized to the electrodes (yellow), is placed into a tube with required PCR reagents, such as dsDNA template (green strands), Taq DNA polymerase (violet) and nucleotides (red). *Step 1.* At the initialization / denaturation temperature the template melts. *Step 2.* In the annealing step the template strands pair with the complementary primers and the polymerase binds to the primer-template hybrid. *Step 3.* The polymerase elongates the immobilized primers. During later cycles of the procedure the extended primer can serve as a template strand for primers attached to the opposite electrode (in the steps 2 and 3). *Step 4.* After denaturation and annealing, the extended primers can form a bridge over the gap of the electrodes via hybridization (lower part) or pair again with the complementary template strands (upper part) (see also Figs. 3(e)-(h)).

## 2. Results and discussion

### 2.1 DEP trapping and immobilization of primers

The first step in the procedure is to prepare a gold nanoelectrode structure on a silicon oxide substrate by electron beam lithography, for trapping the primers to the certain locations on a substrate, i.e. to the selected electrodes. The dimensions of the fabricated electrodes are 20 nm × 20 nm (width × height) for the 8-electrode system (Fig. 1(a)) or 100-170 nm × 20 nm for the fingertip electrodes (Figs. 2 and 3), and the separation between the opposite electrodes is 100-140 nm, roughly corresponding to the length of the grown dsDNA molecule (template for growing: 414 bp partial complementary DNA of chicken avidin[22,23]).

To demonstrate the feasibility of the selective trapping with the multielectrode geometry we used two dye-labelled 40 nt long 5'-hexanethiol-modified oligonucleotides with fluorescent dye molecules, Cy3 and Cy5, attached to the 3'-ends. First, 12 μl of 0.3 μM Cy3-labeled oligonucleotide solution (3 mM Hepes / 2 mM NaOH buffer) was pipetted onto the chip, and the trapping field was created by applying a sinusoidal 1 MHz AC-voltage of 5.0 $V_{pp}$ with DC-offset of 1.3 V to the desired electrode, while keeping the other electrodes grounded. The gathering of the oligonucleotides to the trap, and in particular to the chosen electrode, was studied *in situ* under a confocal-microscope (Olympus FluoView 1000, 60× oil objective; lasers 543 nm and 633 nm were used to excite Cy3 and Cy5 dyes, respectively).

After a successful immobilization of Cy3-modified oligonucleotides, the trapping voltage was switched off and the sample was gently rinsed with distilled water. Second, the same procedure was repeated but now using the opposite electrode to immobilize the Cy5-marked strands. The fluorescence spots of both types of the oligonucleotides could be simultaneously seen on the separate electrodes (see Fig. 1(a)) proving the reliable immobilization and practicability of the technique.

However, the trapping procedure described above is not practical for minute examination of the kinetics of bridging the electrodes by polymerase for the following reasons: *1)* it is a bit laborious method for extending the immobilized primers due to the observed relatively low yield in our (not fully optimized) elongation process (see below), *2)* moreover, the used strands contained 3'-dye molecules and thus were not suitable for being elongated. Due to this reasoning, in order to show that our method of extending the immobilized primers by polymerase to form a dsDNA molecule across the gap is really exploitable, the elongation experiments were carried out using a slightly simplified scheme. As forward and reverse primers we used 5'-thiol-modified 38 nt ssDNA molecules consisting of the $(CT)_8$-spacer followed by the 22 nt long part matching to the terminal sequences of the complementary strands of the 414 bp template.[22,23] The added spacer facilitates the hybridization of the template and attaching of the polymerase to the active/matching part of the immobilized primer. The trapping was carried out similarly as stated above, but now we used an array of adjacent fingertip type electrode pairs (single electrode pairs are presented in Figs. 2 and 3) to gain more data in a single run. Also, both primers were immobilized simultaneously, i.e., ~10 μl of primer solution containing both the primers (~20 nM in the Hepes/NaOH buffer) was pipetted onto the chip and an AC-voltage of 4.5 $V_{pp}$ ($V_{DC} = 0$) was applied between the electrodes for 1-2 minutes. Finally the sample was gently rinsed with 40-50 μl of distilled water and dried with nitrogen flow.

The efficiency of the DEP trapping of the primers was first

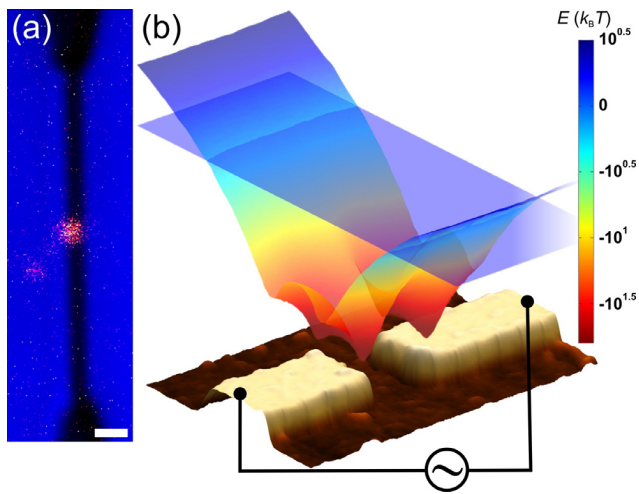

**Fig. 2** (a) A confocal microscope image presenting the trapping and immobilization of 22 nt Cy3-labelled primers in between the fingertip-type electrodes. The scale bar is 500 nm. (b) Simulated DEP trapping potential (color coded surface) in the vicinity of 100 nm wide fingertip-type electrodes (shown as an AFM image). Potential is calculated at the plane just above the electrodes, i.e., about 25 nm above the substrate. Trapping area is the region where the absolute value of the DEP potential is higher than the thermal energy of DNA, i.e., where the potential is below the flat horizontal surface, corresponding to the negative of thermal energy, $-3/2\ k_BT$ ($U_{DEP} + U_{thermal} = 0$). It can be seen that the strongest trapping spots lie at the ends of the electrodes but trapping also takes place along the electrodes.

validated by comparing the finite element method (FEM) simulations of the time-averaged DEP trapping potential ($U_{DEP} = -½\ \alpha\ E^2$, where the estimated value for the polarizability $\alpha \approx 2\times10^{-33}$ Fm$^2$ for a 40 nt ssDNA was used[21]) to the estimated thermal energy of the primers ($U_{thermal} = 3/2\ k_BT$) at the room temperature. The results showed that the absolute value of the DEP potential in the vicinity of the electrodes is well above the thermal energy resulting in a deep trapping well for ssDNA molecules as shown in Fig. 2(b). Yet, to ensure that the primers are trapped and immobilized in a desired way, similar ssDNA molecules labeled with dye molecules (22 nt ssDNA with hexanethiol at the 5'-end and Cy3-modification at the 3'-end) were trapped with the above-mentioned parameters followed by the confocal microscope imaging confirming the result (see Fig. 2(a)). Further, the fluorescence was still visible in the gap region after heating the sample in a PCR buffer to 95 °C, indicating covalent binding of the primers to the electrodes (covalent sulphur-gold bond). Thus, the majority of the trapped primers can be assumed to stay immobilized during the elongation. The results of the trapping of the primers without dye molecules, i.e. the primers feasible for a polymerase extension, were verified by atomic force microscope (AFM) imaging (Veeco, Dimension 3100) (see Figs. 3(a)-(d)).

## 2.2 Extension of primers and bridging the electrodes

As the final step, the chip containing the electrode structure and the immobilized primers (this time two types of primers at each electrode) was placed into a 0.2 ml tube with the reagents and components required for the standard PCR procedure (see Experimental section). The following programme for a thermal cycler (Biometra T3 Thermoblock, Biotron, Germany) was used: *1)* 94 °C, 5 min; *2)* 94 °C, 40 s; *3)* 50 °C, 3 min; *4)* 72 °C, 4 min; *5)* 4 °C, where the cycles 2-4 were repeated 25-50 times. Finally, the chip was removed from the tube and washed with distilled water similarly as after the trapping process.

Since the very low electrical conductivity of a long dsDNA is known to be strongly dependent on the environment (types and concentrations of ions of the buffers),[22,24-26] a reliable electrical verification of successful growth, i.e. bridging the gap, was not practicable. In addition, the utilized single strand spacers in the primers can be considered as insulators. Also, since the low amount of grown individual molecules does not produce strong enough fluorescent signal to employ optical detection either, the successful growth was detected by imaging the samples again with AFM and comparing the images taken before and after growing procedure.

In Fig. 3, examples of the results are shown. In Figs. 3(e) and 3(f) a single ~150 nm long dsDNA molecule, grown during the procedure and comprised of the extended primers attached to the opposite electrodes, is bridging the gap between the electrodes. Figs. 3(g) and 3(h) present alternative results of the process, where elongated primers are paired with template strands via hybridization during the annealing step, forming ~150 nm long dsDNA molecules at the edges of the electrodes.

## 3. Experimental section

### 3.1 Nanoelectrode preparation

The nanoelectrodes were fabricated on an oxidized silicon chip by using electron beam lithography and evaporation of metal (15-18 nm gold on top of 1-2 nm titanium) in an ultrahigh vacuum (UHV) chamber. In addition, PMMA residues from the lift-off were cleaned off the electrode structure with an oxygen plasma flash in a reactive ion etcher. This procedure also made the SiO$_2$ surface hydrophilic, which greatly enhances the DEP-trapping.

### 3.2 DNA strands

The sequences of the thiol-modified (5'end) and either Cy3 or Cy5-dye-labeled (3'end) DNA oligonucleotides used in the optimization procedure of the DEP-trapping were the following: 40 nt long oligos: 5'-(CT)$_{16}$GATGGCTT-3'-Cy3 and 5'-(CT)$_{16}$GAAAAAGC-3'-Cy5; and 22 nt long ssDNA molecules: 5'-GCCAGAAAGTGCTCGCTGACTG-3'-Cy3 (purchased from Biomers as HPLC-purified). In the actual elongation experiments the sequences of the 38 nt forward and reverse primers (without dye-molecules) contained 22 nt long sequences complementary to the template and the additional 16 nt spacer: 5'-(CT)$_8$GCCAGAAAGTGCTCGCTGACTG-3' and 5'-(CT)$_8$TTCTCGACAAGCTTTGCGGGGC-3', where 5' Thiol Modifier C6 S-S (Disulfide) was attached to 5'ends (ordered from Integrated DNA Technologies as dual HPLC-purified).

### 3.3 Reagents for elongation of primers

Reagents for elongation of the primers and their amounts are presented here in the order of mixing (reagent, amount, final concentration): H$_2$O (distilled), 69.9 μl, -; 10× Taq buffer [750

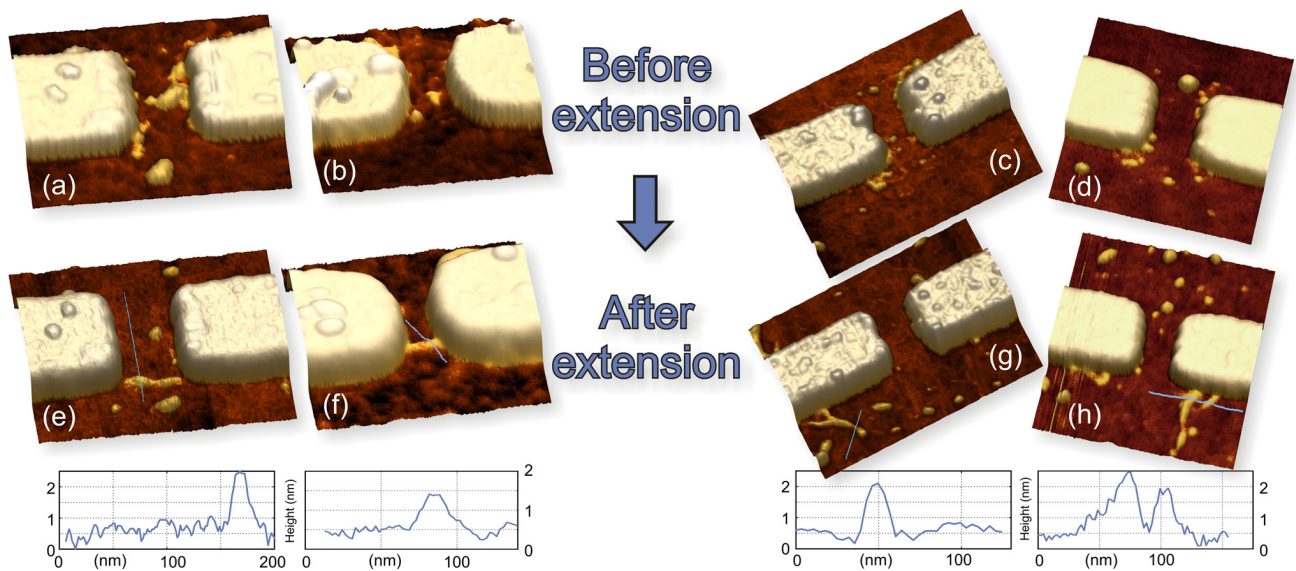

**Fig. 3** (a)-(d) "Before extension" AFM images of bunches of 38 nt primers immobilized to the electrodes. (e)-(h) "After extension" AFM images of the same samples shown in Figs. (a)-(d). Note that the height in all the 3D images is presented in a logarithmic scale. (e)-(f) A single grown dsDNA molecule is bridging the gap between electrodes. (g)-(h) A few dsDNA molecules grown on the edge of the electrode. The non-specifically bound primers have been detached during the elongation. The gap between the electrodes is ~100-120 nm in all samples. The graphs below are cross sections along the blue lines on (e)-(h), showing the characteristic height of the dsDNA molecule on a $SiO_2$ substrate, i.e. 1-2 nm.

nM Tris-HCl (pH 8.8), 200 nM $(NH_4)_2SO_4$, 0.1% Tween 20], 10.0 μl, 1×; $MgCl_2$ (25 mM), 8.0 μl, 2.0 mM; dsDNA template (51 ng/μl), 1.5 μl, 0.76 ng/μl; dNTP mix (2 nM), 10.0 μl, 0.2 mM; Taq polymerase (5 U/μl), 0.6 μl, 1.5 U / 50 μl. Taq DNA polymerase (recombinant) (with Taq buffer and $MgCl_2$) and dNTP mix were purchased from Fermentas.

## 4. Conclusions

In summary, we have demonstrated a novel method to grow individual dsDNA molecules on a chip based on a controllable directing of the primers via dielectrophoresis, and elongation of the immobilized primers by polymerase. The developed technique can serve as a tool in exploiting the on-chip growing of DNA and also PCR on a new level. In this particular proof-of-principle experiment, the yield has been rather low, but there are still several aspects to optimize. By making further improvements on the procedure and designing an electrode pattern suitable for the application in question, it opens up opportunities for detecting single molecules or molecule combinations, and also for fabricating bottom-up based nanostructures from DNA at desired locations on a chip, e.g. sophisticated DNA wiring systems and networks[27] or DNA-based multiswitching units programmable to react on planned targets. Finally, DNA-programming platform can be envisioned, where identical electrode geometries are tailored by using different combinations of primers and template sequences.

## Acknowledgements

We would like to thank J. Ihalainen, J. Ylänne and M. Vihinen-Ranta (Nanoscience Center, University of Jyväskylä) for use of biolab facilities; A. Kuzyk and P. Törmä (Aalto University School of Science and Technology, Helsinki) for momentous discussions. Academy of Finland (projects 218182, 130900 and 115976) is acknowledged for financial support. V.L. thanks Finnish Academy of Science and Letters (Väisälä Foundation), Finnish Cultural Foundation (Central Finland Regional Fund), Finnish Foundation for Technology Promotion (TES), and National Doctoral Programme in Nanoscience (NGS-NANO). J.L. thanks Tampere Graduate Program in Biomedicine and Biotechnology. B.S. thanks Centre of Expertise Programme, The Nanotechnology Cluster Programme 090002-501.

## Notes and references

[a] Nanoscience Center, Department of Physics, University of Jyväskylä, P.O. Box 35, FI-40014, Jyväskylä, Finland, Fax: +358 14 260 4756; Tel: +358 14 260 4722; E-mail: veikko.linko@jyu.fi
[b] Institute of Biomedical Technology, University of Tampere and Tampere University Hospital, FI-33014, Tampere, Finland
[c] Nanoscience Center, Department of Biological and Environmental Science, University of Jyväskylä, P.O. Box 35, FI-40014, Jyväskylä, Finland. Present address: Department of Biochemistry, Erasmus University Medical Center, Dr. Molewaterplein 50, 3015 GE Rotterdam, The Netherlands